\documentclass[preprint,showpacs,preprintnumbers,amsmath,amssymb,nofootinbib]{revtex4}

\usepackage{epsfig}

\newcommand{\be}{\begin{equation}}
\newcommand{\ee}{\end{equation}}
\newcommand{\beqq}{\setlength\arraycolsep{2pt}\begin{eqnarray}}
\newcommand{\eeqq}{\vspace{0cm} \end{eqnarray}}
\newcommand{\bea}{\begin{eqnarray}}
\newcommand{\eea}{\end{eqnarray}}

\begin{document}

\title{Quantum Brownian motion in an analog Friedmann-Robertson-Walker geometry}

\author{C. H. G. \surname{Bessa}}
\email{carlos@cosmos.phy.tufts.edu}
\affiliation{Departamento de F\'{\i}sica, Universidade Federal da Para\'{\i}ba, \\
Caixa Postal 5008, CEP 58051-970, Jo\~ao Pessoa, PB, Brazil}
\author{V. B. \surname{Bezerra}}
\email{valdir@fisica.ufpb.br}
\affiliation{Departamento de F\'{\i}sica, Universidade Federal da Para\'{\i}ba, \\
Caixa Postal 5008, CEP 58051-970, Jo\~ao Pessoa, PB, Brazil}
\author{E. R. \surname{Bezerra de Mello}}
\email{emello@fisica.ufpb.br}
\affiliation{Departamento de F\'{\i}sica, Universidade Federal da Para\'{\i}ba, \\
Caixa Postal 5008, CEP 58051-970, Jo\~ao Pessoa, PB, Brazil}
\author{H. F. \surname{Mota}}
\email{hmota@fisica.ufpb.br}
\affiliation{Departamento de F\'{\i}sica, Universidade Federal da Para\'{\i}ba, \\
Caixa Postal 5008, CEP 58051-970, Jo\~ao Pessoa, PB, Brazil}

\begin{abstract}

In this paper we study the effects of quantum scalar field vacuum fluctuations on scalar test particles in an analog model for the Friedmann-Robertson-Walker spatially flat geometry. In this scenario, the cases with one and two perfectly reflecting plane boundaries are considered as well the case without boundary. We find that the particles can undergo Brownian motion with a nonzero mean squared velocity induced by the quantum vacuum fluctuations due to the time dependent background and the presence of the boundaries. Typical singularities which appears due to the presence of the boundaries in flat spacetime can be naturally regularized for an asymptotically bounded expanding scale function. Thus, shifts in the velocity could be, at least in principle, detectable experimentally. The possibility to implement this observation in an analog cosmological model by the use of a Bose-Einstein condensate is also discussed.   

\end{abstract}

		
\maketitle
\baselineskip=14pt
\section{Introduction}
\label{secint}

The idea that a quantum fluctuating field may induce Brownian motion of test particles  has been a topic of interest in recent years. In general,  these particles are treated as classical point particles that interact with a vacuum fluctuating field and these fluctuations induce the Brownian motion \cite{gs99,jr92,wkf02,yf04,yc04,f05,ycw06,hwl08,sw08,sw09,bbf09,pf11,whl12,lms14,lrs16}. Although for a Minkowski vacuum the motion is unclear, as was noted in Refs. \cite{gs99, jr92}, in flat spacetime such motion appears when one or more reflecting boundaries are present. In this scenario a nonzero  mean squared velocity, that depends on the boundary position is found in the transverse and parallel directions to the boundary \cite{yf04,yc04,lms14,lrs16}. Furthermore, some singularities appears in the mean squared velocity even when the usual renormalization techniques, with respect to the Minkowski background, are implemented \cite{yf04,yc04}.  One of these singularities is due to the boundaries and the other correspond to the duration of a round trip of the signal between the particle position and the boundary location. Although, as was noted in \cite{lms14,lrs16}, they can be regularized when a switching process is considered and a non-instantaneous interaction between the particle and the field is given. 

Another possibility to test particles undergo Brownian motion is when one consider vacuum quantum fluctuations in a time-dependent background such as in a Friedmann-Robertson-Walker (FRW) geometry.  In this case the mean squared velocity may depends on the scale factor of the universe, $a(t)$. In this context, in Ref. \cite{bbf09} different type of couplings and a variety of scale factors were considered for a point charged particle coupled to a fluctuating eletromagnetic field. Two different kinds of interactions were considered. The first one was for particles that move, in the average, on geodesics. The second kind considered concerns particles bounded by an external non-fluctuating force that cancels locally the effects of the expansion. In both cases, the results exhibited a dependence on the redshift and the characteristic spacetime curvature.  

In recent years, a lot of studies using classical or quantum fluids to mimic several aspects of the general relativity theory for cosmological and black hole spacetimes have been considered and in this way simulates some semi-classical gravitational phenomena, such as Hawking radiation and cosmological particle production (see for example Ref. \cite{blv11} and references therein for an extensive review). The basic idea resides in the fact that the linearized perturbation for an inviscid and irrotational fluid describes the same equation of motion of a massless scalar field in curved spacetime. These mentioned phenomena can be observed in principle if we have to our disposal a system such as a Bose-Einstein condensate (BEC) which has been extensively investigated  and seems to constitute an excellent scenario to furnish a possible comprehension of the astrophysical Hawking radiation, particle production in a cosmological scale and superradiance \cite{jwvg07,pfl10,gacz00,gacz01,blv02,blv03,ff03,bm03,bm032,bcl04,ff04,bffrc08,cfrbf08,mp09,bf14,rplw15}.

Using the ideas above, we analyze in this paper the Brownian motion of a test scalar point particle in a fluid analog FRW geometry. To model this geometry we consider an isotropically expanding BEC. Thus, the atoms that constitute the condensate are treated as point particles and they interact with a fluctuating scalar field. This field represents the propagation of acoustic perturbations (i.e. phonons) in a time dependent expanding BEC, analogous to a FRW geometry. In this scenario, we also consider the cases with and without plane boundaries. 
In the presence of one plane or two planes, the Dirichlet boundary conditions are used to derive all results concerning the  velocity dispersions. From these, we find no extra difficulties to obtain the velocity dispertions corresponding to the Neumann boundary conditions.
Thus, we point out what are the differences and similarities between these quantities when Dirichlet and Neumann boundary conditions are used.
One of our general results, in the sense that do not depend on the specific boundary condition used shows how to relate the contributions of the expansion and the boundaries on the particles motion. A second general result, shows that some boundary dependent divergencies that appears in a flat spacetime scenario disappear here, as a consequence of the time dependent expanding background which can simulate a smooth switched function. 

As vacuum fluctuations are strongly anti-correlated it is a physical effect hard to be observed. The Casimir effect and the Lamb shift are two example of this manifestation. If one deals with an external source of energy or a time dependent background this anti-correlations can be upset and the particle undergoes Brownian motion, as was noted in Refs. \cite{bbf09}, \cite{pf11} and \cite{pf14}. So the BEC expanding model considered in this paper can be an operational mean where the effects of a quantum fluctuating field can be measured in a possible experiment. 
 
The paper is organized as follows: in Sect. \ref{secbackground} we review some basic aspects of an analog FRW geometry using BEC. In particular, how one can build an effective expanding analog cosmological metric from the condensate's fundamental equation, the Gross-Pitaevskii equation. In Sect. \ref{secscalar}, we study the behavior of a scalar particle in curved spacetime and applies the results to the equation of motion for this particle in a 4-dimensional curved spacetime. Finally in Sect. \ref{secanalog} we study the Brownian motion properly. We will see how the fluctuating field in an analog FRW geometrical scenario modifies the average mean squared velocity for free and bound particles. The case of one and two plane boundaries are also investigated. In Sect. \ref{secconclu} we discuss the main results as well as their possible experimental measurement.

\section{The Background: Analog FRW spacetime}\label{secbackground}

A spatially flat Friedmann-Robertson-Walker (FRW) universe with metric

\begin{equation}\label{frw1}
ds^2_{FRW} = -c^2dt^2 + a^2(t)(dx^2 + dy^2 + dz^2),
\end{equation}
represents a homogeneous and isotopic universe with a time dependent scale factor $a(t)$ and light velocity $c$. This metric admits a conformal transformation in the time variable of the type, $dt =a(\eta)d\eta$, where $\eta$ is the conformal time, in which case it can be written as

\begin{equation}\label{frw2}
ds^2_{FRW} = a^2(\eta)(-c^2d\eta^2 + dx^2 + dy^2 + dz^2).
\end{equation}
This metric is conformally related with the Minkowski metric with a conformal factor $a^2(\eta)$.

In this paper we are interested to study the Brownian motion of scalar particles in an analog FRW-spacetime, in which case a Bose-Einstein Condensate (BEC) is one of the most promising system to generate this kind of analog models. In order to review some aspects of how to generate this type of analog model using BEC's, let us consider a field operator $\psi(t, {\bf x})$ with equation of motion given by

\begin{equation}\label{eqbec0}
i\hbar\frac{\partial \psi}{\partial t} = \left[-\frac{\hbar^2}{2m}\nabla^2 + V_{ext} + U\psi^{\dagger}\psi\right]\psi,
\end{equation}
where $\psi^\dagger$ and $\psi$ represents the creator and annihilator boson operators, which obey the usual commutation relations

\begin{equation}
\left[\psi(t, {\bf x}), \psi(t, {\bf x}')\right] = \left[\psi^\dagger(t, {\bf x}), \psi^\dagger(t, {\bf x}')\right] = 0; \:\:\: \left[\psi(t, {\bf x}), \psi^\dagger(t, {\bf x'})\right] = \delta\left( {\bf x} - {\bf x}' \right).
\end{equation}
In Eq. (\ref{eqbec0}), $V_{ext}$ represents any arbitrary external potential and $U = 4\pi\hbar^2L/m$ is the potential interaction between two particles in the condensate, with $L$ being the scattering length and $m$ the mass of the atoms. To establish an analogy with the cosmological spacetime, the field above is expanded as follows,

\begin{equation}
\psi(t, {\bf x}) = \psi_0(t, {\bf x}) + \delta\varphi(t, {\bf x})
\end{equation} 
where $\psi_0(t, {\bf x})$ is a classical field associated with the condensate and $\delta\varphi(t, {\bf x})$ corresponds to a small quantum or thermal perturbation fluctuation, such that $\langle\delta\varphi\rangle = 0$. Thus, $\langle\psi\rangle = \psi_0$, and a Gross-Pitaevskii equation (GPE) is obtained, which describes the condensation in terms of the classical field $\psi_0$ and is given by

\begin{equation}\label{eqGP}
i\hbar\frac{\partial \psi_0}{\partial t} = \left[-\frac{\hbar^2}{2m}\nabla^2 + V_{ext} + U|\psi_0|^2\right]\psi_0.
\end{equation}

It is possible to show that $\psi_0(t, {\bf x})$ can be written in terms of a phase, $\phi(t, {\bf x}$), and a real density, $n(t, {\bf x})$ as

\begin{equation}
\psi_0(t, {\bf x}) = \sqrt{n(t, {\bf x})}e^{i\phi(t, {\bf x})}.
\end{equation}
If we consider a linearization such that: $ n \approx n_0 + n_1$, $\phi \approx \phi_0 + \phi_1$, then Eq. (\ref{eqGP}) turns into  a field equation for $\phi_1$ similar to the equation of motion of a massless scalar field in curved space, as follows

\begin{equation}
\frac{1}{\sqrt{-g}}\partial_\mu(\sqrt{-g}g^{\mu\nu}\partial_\nu\phi_1) = 0,
\end{equation} 
where

\[
g_{\mu\nu}= \left(\frac{n_0}{c_s}\right)
  \begin{bmatrix}
    -(c_s^2 - v^2) & -v_j \\
    -v_j & \delta_{ij} 
  \end{bmatrix}
\]
with $c_s^2 = Un_0/m$ being the speed of sound in the condensate and ${\bf v} = \hbar/m\nabla\phi_0$ the background velocity which can be written in terms of the irrotational field. When the background flow is zero $({\bf v} = 0)$, we get

\[
g_{\mu\nu}= \left(\frac{n_0}{c_s}\right)
  \begin{bmatrix}
    -c_s^2 & 0 \\
    0 & \delta_{ij} 
  \end{bmatrix}
\]
which represents an effective FRW metric and can be written in the form:

\begin{equation}\label{eff1}
d\bar{s}^2_{eff} = \Omega_0^2\left[-c_0^2b^{1/2}(t)dt^2 + b^{-1/2}(t)(dx^2 + dy^2 + dz^2)\right],
\end{equation}
where $\Omega_0^2 = n_0/c_0$ and $b(t)$ is a time dependent scaling function given in terms of the sound velocity as 

\begin{equation}\label{eqb1}
b(t) = \left(\frac{c_s(t)}{c_0}\right)^2,
\end{equation}
with $c_0$ being  the constant sound velocity. Note that we can take, without loss of generality, $b(t_0) = 1$ and thus the time dependent sound velocity, $c_s(t)$, should be defined as

\begin{equation}\label{eqvelocity}
c_s^2(t) = \frac{4\pi\hbar^2}{m^2}n_0L(t).
\end{equation}
Here we have considered a time-dependent scattering length although the time dependence could be introduced in any of the variables of Eq. (\ref{eqvelocity}). For instance, in Ref. \cite{pfl10} the authors considered a time dependent density $n_0(t)$. Note that from Eq. (\ref{eqb1}) we conclude that when $b(t)$ decreases or increases with time we have an analog of an expanding or contracting universe, respectively. 

We can also change Eq. (\ref{eff1}) to a conformal time, $\eta$, defined in terms of the coordinate time by the relation $d\eta = \sqrt{b(t)}dt$. In this case, the metric given by Eq. (\ref{eff1}) turns into

\begin{equation}\label{eff2}
d\bar{s}^2_{eff} = \bar{a}^2_{eff}(\eta)(-c_0^2d\eta^2 + dx^2 + dy^2 + dz^2),
\end{equation}
where $\bar{a}^2_{eff}(\eta) = \Omega_0^2b^{-1/2}(\eta)$.

Note that in the metric given by Eq. (\ref{eff2}), $\Omega_0$ is a dimensional constant, so that the metric $d\bar{s}_{eff}$ has different dimension of the coordinates $(\eta, x, y, z)$. In order to have a world line element with the same dimension of these coordinates it is convenient to redefine a new line element using the parametrization $ds_{eff}^2 = \Omega_0^{-2}d\bar{s}^2_{eff}$. Thus, the metric (\ref{eff2}) turns into

\begin{equation}\label{eff3}
ds^2_{eff} = a^2_{eff}(\eta)(-c_0^2d\eta^2 + dx^2 + dy^2 + dz^2),
\end{equation}
where

\begin{equation}\label{eqaeff77}
a_{eff}(\eta) = b^{-1}(\eta) = \left(\frac{c_0}{c_s(t)}\right)^2.  
\end{equation}
is a dimensionless effective scale factor with $c_s(t)$ given by Eq. (\ref{eqvelocity}). 

In Section \ref{secanalog} we will represent the correlation function for the scalar particle velocity in the fluid in the geometry given by Eq. (\ref{eff3}), which is equivalent to the one given by Eq. (\ref{frw2}).

\section{Scalar particle motion in curved spacetime}\label{secscalar}

In this section we review some aspects of the motion of a scalar particle in a curved spacetime. Firstly, let us consider a scalar point particle with mass $m$ and charge $q$ interacting with a massless scalar field $\phi$ in an arbitrary curved spacetime. The equation of motion for this case is given by \cite{ppv11}

\begin{equation}\label{eqmotion1}
m\frac{Du^{\mu}}{d\tau} = q \left(g^{\mu\nu} + u^{\mu}u^{\nu}\right)\nabla_{\nu}\phi.
\end{equation}
Here $D/d\tau$ is the covariant derivative, $u^{\mu}$ is the particle's 4-velocity and $\tau$ is its proper time. The relation above says that the scalar field applies on the particle a force similar to the Lorentz force. 

One consequence that a scalar field has zero spin is that the mass associated to the particle varies and satisfy an equation of the kind

\begin{equation}\label{mass1}
\frac{dm}{d\tau} = -qu^\mu\nabla_\mu\phi.
\end{equation}
 
The field equation associated to a massless scalar field ($\phi(x)$) is, in general, given by

\begin{equation}\label{wave1}
\left[\frac{1}{\sqrt{-g}}\partial_\alpha(\sqrt{-g}g^{\alpha\beta}\partial_\beta) + \xi R\right]\phi(x) = -4\pi\mu(x),
\end{equation}
where $g$ is the determinant of the  metric $g_{\mu\nu}$, $R$ is the Ricci scalar, $\xi$ an arbitrary curvature coupling and $\mu(x)$ the scalar charge density which is given in terms of a Dirac delta function \cite{bd82}. The retarded solution associated to the wave equation (\ref{wave1}) is given by

\begin{equation}\label{field1}
\phi(x(\tau)) = q \int_\gamma G_{ret} (x(\tau), x'(\tau')) d\tau',
\end{equation} 
where $G_{ret}$ is the retarded Green function associated to the particle's world line $\gamma$. 

According to Refs. \cite{ppv11} and \cite{q00}, Eq. (\ref{mass1}) can be rewritten in terms of the gradient of Eq. (\ref{field1}). So the mass variation in a 4-dimensional spacetime is given by

\begin{equation}\label{mvariation2}
\frac{dm}{d\tau} = - \frac{1}{12}(1 - 6\xi)q^2R - q^2u^\mu\int_{-\infty}^{\tau''}\nabla_\mu G_{ret}(x(\tau), x(\tau'))d\tau'.
\end{equation}

Note that the first term of the equation above vanishes when $R = 0$ (flat spacetime) or $\xi = 1/6$, for a conformal 4-dimensional spacetime, even if $R \neq 0$. Otherwise, it can be shown that for a flat or conformal flat spacetimes, the integral above is always zero. Thus under these conditions both terms in the right-hand side of Eq. (\ref{mvariation2}) vanishes. Therefore the mass $m$, associated to a scalar point particle is constant for both cases, flat and conformal flat spacetimes. Under these conditions,  Eq. (\ref{mass1}) is equal to zero and the equation of motion (\ref{eqmotion1}) simplifies to   

\begin{equation}\label{eqmotion2}
m\frac{Du^{\mu}}{d\tau} = q g^{\mu\nu} \nabla_{\nu}\phi.
\end{equation}

The expression above represents the equation of motion for a scalar point particle moving in flat or conformal flat spacetime and is similar to a Lorentz force law for eletromagnetism. In the next section we will associate Eq. (\ref{eqmotion2}) with a fluctuating force that drives the Brownian motion for scalar particles in the fluid analog spacetime represented by metric (\ref{eff3}). This force is given in terms of a fluctuating massless scalar field which in turns describes the fluid.    

\section{Brownian motion of scalar particles in analog FRW geometry }\label{secanalog}

In this section we will first develop the formalism for the movement of a scalar point particle in a FRW spacetime making use of the results of Secs. \ref{secbackground}, \ref{secscalar} and Ref. \cite{bbf09}. Because of the similarity between  Eqs. (\ref{frw2}) and (\ref{eff3}), the results can be applied to an analog FRW geometry.   

We found in previous section that the equation of motion for a scalar point particle in a curved spacetime which is given by Eq. (\ref{eqmotion2}). Now, let us write this equation in the form:

\begin{equation}
 m\frac{D u^\mu}{d\tau} = f^\mu,
\end{equation}
Using the fact that the FRW metric is homogeneous and isotropic we can choose a particular $i$-direction to the particle motion. The force term $f^i$, can be replaced by a fluctuating quantized  term plus a classical and non-fluctuating external term ($f^i_{ext}$), in such a way that $f^i \rightarrow f^i + f^i_{ext}$. In what follows, we will consider the external term $f^i_{ext} = 0$ or more conveniently $f^i_{ext} = 2m\frac{\dot{a}}{a}u^i$. In the latter case we say that the particles are bounded due to the influence of the external force. The visible effect is that $f_{ext}$ cancels locally the effects of the expansion. Another effects will be clarified in the next section. In the former case, when $f_{ext} = 0$ we say that the particles are free and they move apart as the system expands. 

Making use of the results of Section 3 of Ref. \cite{bbf09} we can obtain the following equation

\begin{equation}\label{eqfree1}
\frac{1}{a^2}\frac{d}{dt}\left(a^2u^i\right) = \frac{f^i}{m},
\end{equation}
to the case where $f_{ext} = 0$ (the particles move freely) and 

\begin{equation}\label{eqdudt}
\frac{du^i}{dt} = \frac{f^i}{m},
\end{equation}
to the case $f^i_{ext} = 2m\frac{\dot{a}}{a}u^i$ (the particles are bounded). We will analyze both cases in Sects. \ref{secfrw} and \ref{secbound}.


\subsection{ Brownian motion for free  particles in an analog FRW geometry}\label{secfrw}



Integrating the Eq. (\ref{eqfree1}) and considering that the particle is at rest in an initial time $t_0$, $(u^i(t_0) = 0)$, we get the correlation function for the velocity in terms of a fluctuating force $f^i$:


\begin{equation}\label{eqfluctuation1}
\langle u^i(t_f, r_1)u^i(t_f, r_2)\rangle = \frac{1}{m^2a^4(t_f)}\int\int dt_1dt_2a^2(t_1)a^2(t_2)\langle f^i(t_1,r_1)f^i(t_2,r_2)\rangle_{FRW},
\end{equation}
where the subscript $FRW$ denotes a vacuum fluctuating function in the Friedmann-Robertson-Walker geometry. We have also considered that the fluctuating force has the following properties: $\langle f^i(t_1,r_1)f^i(t_2,r_2)\rangle_{FRW} \neq 0$ and $\langle f^i(t_1,r_1)\rangle_{FRW} = 0$.

Using Eq. (\ref{eqmotion2}) for the fluctuating force in the geometry given by metric (\ref{frw1}), we obtain

\begin{equation}\label{eqforce21}
f^i(t, r) = qg^{ii}\nabla_i\phi = qa^{-2}(t)\partial_i\phi.
\end{equation}

Substituing Eq. (\ref{eqforce21}) into Eq. (\ref{eqfluctuation1}), we find

\begin{equation}\label{eqmain0}
\langle u^i(t_1, r_1)u^i(t_2, r_2)\rangle = \frac{q^2}{m^2a^4_f}\partial_{i_1}\partial_{i_2}\int\int dt_1dt_2\langle\phi(t_1, r_1)\phi(t_2, r_2)\rangle_{FRW}.
\end{equation}

Now, let us consider the fact that the correlation function for a massless scalar field in the conformal 4-dimensional FRW-spacetime is related to Minkowski (flat) spacetime by \cite{bd82},

\begin{equation}\label{eqconforme1}
\langle\phi(\eta_1, r_1)\phi(\eta_2, r_2)\rangle_{FRW} = a^{-1}(\eta_1)a^{-1}(\eta_2)\langle\phi(\eta_1, r_1)\phi(\eta_2, r_2)\rangle_{flat},
\end{equation}
where the subscript in the right-hand side denotes that the vacuum fluctuation is taken in flat spacetime.

Writing Eq. (\ref{eqmain0}) in terms of the conformal time $(dt = ad\eta)$  and taking into account Eq. (\ref{eqconforme1}) we obtain

\begin{equation}\label{eqmain1}
\langle u^i(t_1, r_1)u^i(t_2, r_2)\rangle = \frac{q^2}{m^2a_f^4}\partial_{i_1}\partial_{i_2}\int\int d\eta_1 d\eta_2\langle\phi(\eta_1, r_1)\phi(\eta_2, r_2)\rangle_{flat}.
\end{equation}

Note that the equation above refers to the dispersion in the coordinate velocity. In FRW geometry under consideration the proper distance is defined in terms of the coordinate distance, $r$, by $l_f = a_fr$, at a time $\eta = \eta_f$ and as a consequence, the proper velocity of the particles is $v = a_fu$. Thus, Eq. (\ref{eqmain1}) in terms of $v$, is written as

\begin{equation}\label{eqproperv1}
\langle v^i(t_1, r_1)v^i(t_2, r_2)\rangle = \frac{q^2}{m^2a_f^2}\partial_{i_1}\partial_{i_2}\int\int d\eta_1 d\eta_2\langle\phi(\eta_1, r_1)\phi(\eta_2, r_2)\rangle_{flat}.
\end{equation}

These integrals are the same integrals that appear in flat spacetime, which are divergent. In fact, in flat spacetime the scale factor is equal to one and the conformal time  is equal to the coordinate time. So, in order to obtain a finite and well defined result we adopt some renormalization procedure. We shall apply it by subtracting the corresponding integral in Minkowski spacetime. Therefore in  the case where the correlation function is expressed by the renormalized Minkowski correlation function, the result is zero. In this case, there is no Brownian motion for free particles when using  Eq. (\ref{eqforce21}). One possible interpretation for this result is that the expansions prevents the Brownian motion once such particles moves apart one from the other. We will see in the Sect. \ref{secbound} that this does not happens when the particles are bounded by an external non-fluctuating force. However, first we will study in Sects. \ref{onep1} and \ref{twop1}, the interesting case of a non-null renormalized correlation function which applies when boundary planes are present.   

\subsubsection{First case: one plane boundary}\label{onep1}

In this section we investigate the effects of one perfectly reflecting plane boundary, located at $z = 0$, on the massless scalar field correlation function.  We will see that typical singularities that appears in flat spacetime also appears here once the scale factor is not present in the integrand of Eq. (\ref{eqproperv1}). Note that, so far in this section, we have developed the general formalism for a FRW geometry given by metric (\ref{frw1}) or (\ref{frw2}). Now we will apply this formalism for the analog FRW geometry. Writing Eq. (\ref{eqproperv1}) in terms of metric (\ref{eff3}) one needs to replace $a$ by $a_{eff}$ at a certain final time $\eta = \eta_f$. However, as the scale factor is a constant in Eq. (\ref{eqproperv1}), we do not need to worry about its form by now. We will specify $a_{eff}$ in Section \ref{secbound} when its form will be relevant to the integrals of the velocity correlation function. The form of $a_{eff}$ gives the behavior of the condensate expansion. In this situation, depending on the choice of $a_{eff}$, the singularities cited previously disappear.  

In what follows, let us consider  Eq. (\ref{eqproperv1}) with the correlation function associated with a massless scalar field with a plane boundary located at $z = 0$ in flat spacetime and taking into account Dirichlet boundary conditions \cite{bd82}. Thus, we get  

\begin{equation}
\label{eqboundary1}
\langle\phi(\eta_1, r_1)\phi(\eta_2, r_2)\rangle_{flat} = \frac{1}{4\pi^2}\left[\frac{1}{-c_0^2(\eta_1 - \eta_2)^2 + r^2} - \frac{1}{-c_0^2(\eta_1 - \eta_2)^2 + {\tilde{r}}^2}\right],
\end{equation}
where $c_0$ is the speed of sound in the condensate at a certain initial time and spatial separation $r$, $\tilde{r}$  given by

\begin{equation}\label{eqr}
r^2 = \sum_{i=1}^{3}(\Delta x^i)^2 = \Delta x^2 + \Delta y^2 + \Delta z^2,
\end{equation}
and

\begin{equation}\label{eqrtilde}
\tilde{r}^2 = \Delta x^2 + \Delta y^2 + \Delta \tilde{z}^2,
\end{equation}
with $(\Delta x^i)^2 = (x_1^i - x_2^i)^2$ and $\Delta \tilde{z}^2 = (z_1 + z_2)^2$.

 The right hand side of Eq. (\ref{eqboundary1}) has a pure Minkowski boundarless term, given by the first term, and the second one is the contribution due the single plane. This term is finite in the coincidence limit. 

By substituting Eq. (\ref{eqboundary1}) into (\ref{eqproperv1}), a divergent result is obtained in the coincidence limit. This divergence comes from the pure Minkowski contribution of the correlation function. The renormalized expression is given by subtracting the Minkowski term in Eq. (\ref{eqboundary1}). After that, we  can differentiate Eq. (\ref{eqproperv1}) with respect to coordinate $z$ for a dispersion perpendicular to the plane and with respect to $x (y)$ for a dispersion parallel to the plane. They are shown in the following expressions,

\begin{eqnarray}\label{noscalex1}
\langle v^x(\eta_1, r_1)v^x(\eta_2, r_2)\rangle = \frac{-q^2a_f^2}{2\pi^2m^2}\int_{-\infty}^{+\infty}\int_{-\infty}^{+\infty} d\eta_1 d\eta_2 \{f_2(\eta, \tilde{r})  + 4\Delta x^2f_3(\eta, \tilde{r})\},
\end{eqnarray}

\begin{eqnarray}\label{noscaley1}
\langle v^y(\eta_1, r_1)v^y(\eta_2, r_2)\rangle = \frac{-q^2a_f^2}{2\pi^2m^2}\int_{-\infty}^{+\infty}\int_{-\infty}^{+\infty} d\eta_1 d\eta_2 \{f_2(\eta, \tilde{r})  + 4\Delta y^2f_3(\eta, \tilde{r})\}
\end{eqnarray}
and

\begin{eqnarray}\label{noscalez1}
\langle v^z(\eta_1, r_1)v^z(\eta_2, r_2)\rangle = \frac{q^2a_f^2}{2\pi^2m^2}\int_{-\infty}^{+\infty}\int_{-\infty}^{+\infty} d\eta_1 d\eta_2 \{f_2(\eta, \tilde{r})  + 4\Delta {\tilde{z}}^2f_3(\eta, \tilde{r})\},
\end{eqnarray}
where the function $f_n(\eta, z)$ in the equations above is given by

\begin{eqnarray}\label{generalfn}
f_n(\eta, z) = \frac{1}{[c_0^2(\eta_1-\eta_2)^2 - {\tilde{r}}^2]^n}.
\end{eqnarray}
Note that, as the plane is placed in the $z$ direction, $\Delta x = \Delta y = 0$, in the coincidence limit and the dispersion in the $x$ and $y$ direction are always the same.  

The respective dispersion  after integrating Eqs.  (\ref{noscalex1}), (\ref{noscaley1}) and (\ref{noscalez1}), in the coincidence limit, is given by

\begin{eqnarray}\label{eqfreex1}
\langle (\Delta v^x)^2\rangle = \langle (\Delta v^y)^2\rangle = \frac{-q^2\eta}{64\pi^2m^2a_f^2z^3c_0}\ln\left[\left(\frac{c_0\eta+2z}{c_0\eta-2z}\right)^2\right],
\end{eqnarray}
which represents the quantum dispersion in the velocity along the parallel direction to the planes and

\begin{eqnarray}\label{eqfreez1}
\langle (\Delta v^z)^2\rangle = \frac{q^2\eta}{32\pi^2m^2a_f^2z^3c_0}\left\{ \frac{4c_0\eta z}{(c_0^2\eta^2-4z^2)} - \ln\left[\left(\frac{c_0\eta+2z}{c_0\eta-2z}\right)^2\right]    \right\},
\end{eqnarray}
which represents the quantum dispersion in the velocity along the perpendicular direction to the planes.

Equations (\ref{eqfreex1}) and (\ref{eqfreez1}) shows that some singularities are present, one at the plane location $z = 0$ and the other at $c_0\eta = 2z$, which represents a time interval for a round trip from the particle location to the plane. This result is similar to the one found for a charged test particle under the influence of a vacuum fluctuating electric field close to a single plane in flat spacetime \cite{yf04}.

In the limit of long time $\eta \gg z/c_0$, Eqs. (\ref{eqfreex1}) and (\ref{eqfreez1}) become, respectively,
\begin{equation}
\label{eqfreex11}
\langle (\Delta v^x)^2\rangle \approx \frac{-q^2\eta}{64\pi^2m^2a_f^2z^3c_0}\left\{\frac{8z}{c_0\eta} + \frac{32z^3}{3c_0^3\eta^3}    \right\} \approx \frac{-q^2}{8\pi^2m^2a_f^2c_0^2z^2} 
\end{equation}
and

\begin{equation}
\label{eqfreez11}
\langle (\Delta v^z)^2\rangle \approx \frac{q^2\eta}{32\pi^2m^2a_f^2z^3c_0}\left\{- \frac{4z}{c_0\eta} + \frac{16z^3}{3c_0^3\eta^3}    \right\} \approx \frac{-q^2}{8\pi^2m^2a_f^2c_0^2z^2} 
\end{equation}
where the anisotropy present in Eqs. (\ref{eqfreez1}) and (\ref{eqfreex1}) is broken in the long time regime. A different result compared to the eletromagnetic case.

Now, if we taken into account the Neumann boundary condition, namely, $\partial_z\phi=0$ at $z=0$, the expression for the two-point function can be obtained from \eqref{eqboundary1} by changing the sign of the second term in the right hand side.  As a consequence of this changing, the expressions for the dispersions of the velocity, given by \eqref{eqfreez1} and \eqref{eqfreex1}, present opposite sign.

\subsubsection{Second  case: two plane boundaries}\label{twop1}

\begin{figure}[htbp]
	\centering
		\includegraphics[scale=0.4]{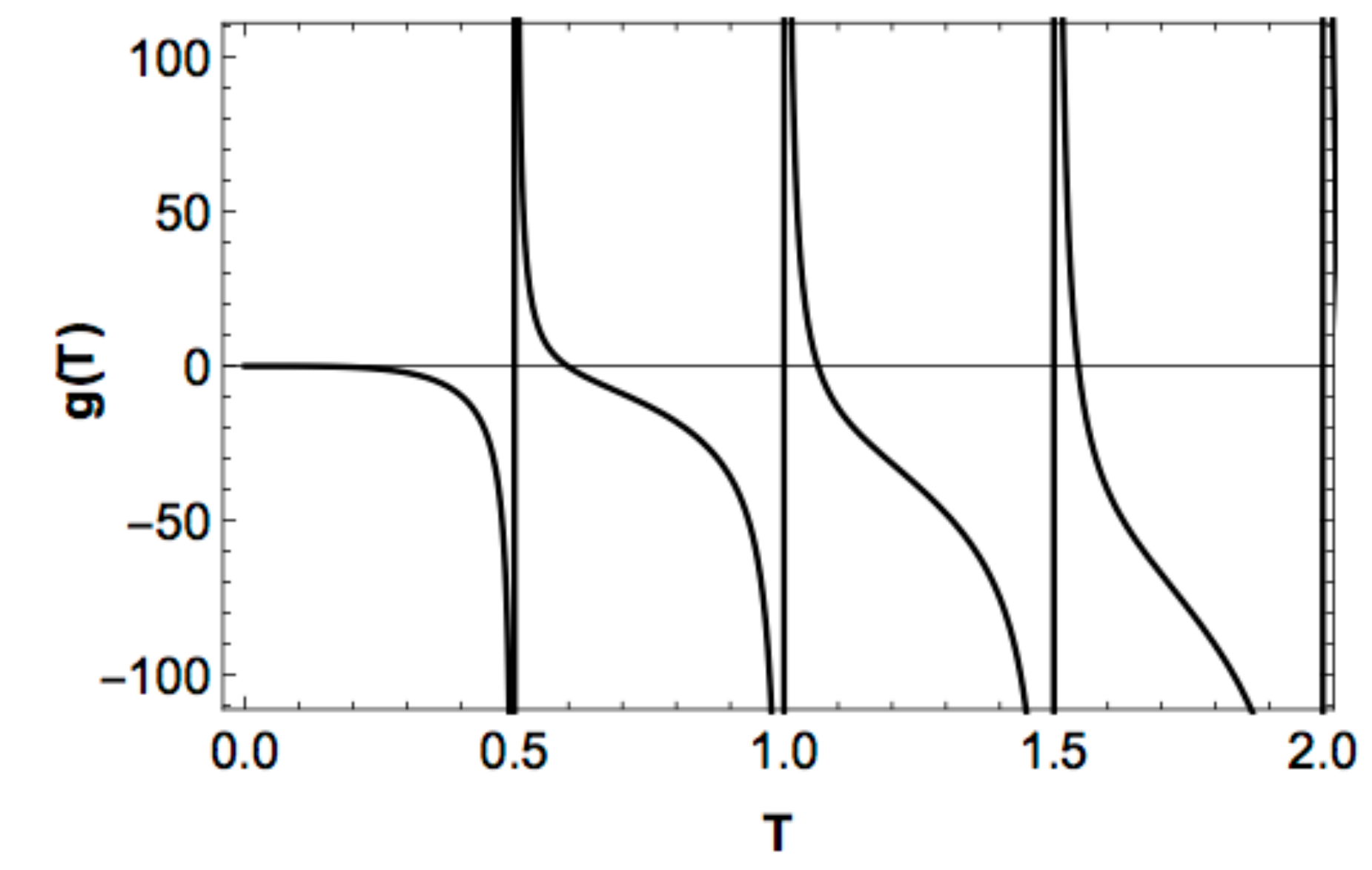}
		\caption{{\it This figure exibit the behavior  of $g(T) \equiv \langle(\Delta v^z)^2\rangle d^3c_0m^2a_f^2/\eta q^2$ as a function of $T=c_0\eta/2d$. It shows the divergences of the velocity dispersion for a time interval corresponding to a round trip of the particles location from the plane at $z = 0$ to the one at $z = d$. This result is similar  to the flat spacetime case.}}
	\label{figfree}
\end{figure}

Now we want to investigate the effect of two parallel plane boundaries. The correlation function associated with a massless scalar field when two perfectly reflecting planes are present,  located at $z = 0$ and $z = d$, taking into account the Dirichlet boundary conditions in Minkowski spacetime, is given by \cite{bd82} 

\begin{eqnarray}\label{eqboundary2}
\langle\phi(\eta_1, r_1)\phi(\eta_2, r_2)\rangle_{flat} = \frac{1}{4\pi^2}\left\{\left[\frac{1}{-c_0^2(\eta_1 - \eta_2)^2 + r^2} \right] +   {\sum}_{l = -\infty}^{'\infty} \left[\frac{1}{-c_0^2(\eta_1 - \eta_2)^2 + r_l^2}\right]  \right. \\ \nonumber \left.- \sum_{l = -\infty}^{\infty} \left[\frac{1}{-c_0^2(\eta_1 - \eta_2)^2 + {\tilde{r}_l}^2}\right]\right\},
\end{eqnarray}
where the prime in the first sum indicates that $l = 0$ is not included. The quantities $r_l$ and $\tilde{r}_l$ are given by 

\begin{equation}\label{eqrl}
r_l^2 = \Delta x^2 + \Delta y^2 + \Delta z_l^2,
\end{equation}
and

\begin{equation}\label{eqrltilde}
\tilde{r}_l^2 = \Delta x^2 + \Delta y^2 + \Delta \tilde{z}_l^2, 
\end{equation}
with $\Delta z_l^2 = (z_1 - z_2 + 2ld)^2$ and $\Delta \tilde{z}_l^2 = (z_1 + z_2 + 2ld)^2$.  The first term in the right hand side of Eq. (\ref{eqboundary2}) is the Minkowski correlation function. The effects due the presence of boundaries are given by the second and third terms.

The dispersion in Eq. (\ref{eqproperv1}) is calculated with the renormalized correlation function for the two planes, which is obtained after excluding the first term in Eq. (\ref{eqboundary2}). Thus, taking the derivatives, we obtain the following result in the coincidence limit in the transverse direction 

\begin{eqnarray}\label{eqfreez2}
\langle (\Delta v^z)^2\rangle = \frac{q^2\eta}{32m^2\pi^2a_f^2d^3c_0}\left\{{\sum}_{l=-\infty}^{'\infty}\left[ -\frac{1}{l^3}\ln\left[\left(\frac{T+l}{T-l}\right)^2\right]  + \frac{2T }{l^2(T^2-l^2)}\right] + \right. \\ \nonumber \left. {\sum}_{l=-\infty}^\infty\left[ -\frac{1}{(Q+l)^3}\ln\left[\left(\frac{T+Q+l}{T-Q-l}\right)^2\right]  + \frac{2T}{(Q+l)^2(T^2-(Q+l)^2)}\right] \right\},
\end{eqnarray}
where $Q = z/d$ and $T = c_0\eta/2d$, and in the parallel direction

\begin{eqnarray}\label{eqfreex2}
\langle (\Delta v^x)^2\rangle = \frac{q^2\eta}{64\pi^2m^2a_f^2d^3c_0}\left\{{\sum}_{l=-\infty}^{'\infty} \frac{1}{l^3}\ln\left[\left(\frac{T+l}{T-l}\right)^2\right] -\right. \\  \nonumber \left. {\sum}_{l=-\infty}^\infty \frac{1}{(Q+l)^3}\ln\left[\left(\frac{T+Q+l}{T-Q-l}\right)^2\right] \right\}.
\end{eqnarray}

In Fig. \ref{figfree} we exhibit the behavior of $\langle (\Delta v^z)^2\rangle$ in units of $d^3c_0m^2a_f^2/\eta q^2$ as a function of $Q$. We can see that there are singular points at the planes location $z = 0$ ($Q = 0$) and $z = d$ ($Q = 1$). These singular points come from the term $Q + l$ present in the denominator of Eqs. (\ref{eqfreez2}) and (\ref{eqfreex2}) when $l = 0$ and $l = -1$. Besides these singularities there are divergences due to a time interval equivalent to a round trip when $T = \pm l$ and $T = \pm Q \pm l$. Note that when $Q = 0$ the last case is identical to the former case.  Because this time depends of $l$ we can have a multiple round trip interval. The same conclusions occur with respect to the $x(y)$-directions. All these singularities are also present in flat spacetime. So, the results obtained in this section for free particles are very similar to the ones found in flat spacetime case. In next section we will study the case of bounded particles. We will see that all singularities that appear here can be naturally removed.   

Once again, let us briefly analyze the modifications in our result if we had used the Neumann boundary conditions at $z=0$ and $z=d$. In this case, the two-point function associated with massless scalar field can be obtained from \eqref{eqboundary2} by changing the sign of the  third term in its right hand side. Thus, the corresponding dispersion relations can be obtained from \eqref{eqfreez2} and \eqref{eqfreex2} by changing the sign of their second term.

\subsection{ Brownian motion for bounded  particles in an analog FRW geometry}\label{secbound}

In this section we begin by considering Eq. (\ref{eqdudt}) where an external non-fluctuating force of the kind $f_{ext}^i = 2mu^i\dot{a}/a$ was considered. This is the equation of motion of scalar point particles. It is worth calling attention to the fact that  the particles do not feel the effects of the expansion. This could be a consequence of the interaction between them which cancels off locally the effect of the expansion. They are named bounded particles. In this model, integrating Eq. (\ref{eqdudt}) with a null initial velocity, we get


\begin{equation}
u^i(t_f) = \frac{1}{m}\int f^i(t, r)dt,
\end{equation}
where $t_f$ is the final time and the correlation function for the coordinate velocity is  

\begin{equation}\label{eqfluf}
\langle u^i(t_f, r_1)u^i(t_f, r_2)\rangle = \frac{1}{m^2}\int\int dt_1dt_2\langle f^i(t_1,r_1)f^i(t_2,r_2)\rangle_{FRW}.
\end{equation}
Let us consider  Eq. (\ref{eqforce21}) as a fluctuating force of the type, $\langle f^i(t_1,r_1)f^i(t_2,r_2)\rangle_{FRW} \neq 0$ and $\langle f^i(t,r)\rangle_{FRW} = 0$.  Them, Eq. (\ref{eqfluf}) reads

\begin{equation}
\langle u^i(t_1, r_1)u^i(t_2, r_2)\rangle = \frac{q^2}{m^2}\partial_{i_1}\partial_{i_2}\int\int dt_1dt_2a^{-2}(t_1)a^{-2}(t_2)\langle\phi(t_1, r_1)\phi(t_2, r_2)\rangle_{FRW}.
\end{equation}

When equation above is written in terms of the conformal time, $dt = a(\eta)d\eta$, and in terms of the massless two point function given by the relation presented in Eq. (\ref{eqconforme1}), it provides the following expression for the velocity dispersion of the bounded particles

\begin{equation}\label{eqmain2}
\langle u^i(\eta_1, r_1)u^i(\eta_2, r_2)\rangle = \frac{q^2}{m^2}\partial_{i_1}\partial_{i_2}\int\int d\eta_1 d\eta_2 a^{-2}(\eta_1)a^{-2}(\eta_2)\langle\phi(\eta_1, r_1)\phi(\eta_2, r_2)\rangle_{flat}.
\end{equation}
In this case,  the proper velocity is given by

\begin{equation}\label{eqproperv2}
\langle v^i(\eta_1, r_1)v^i(\eta_2, r_2)\rangle = \frac{q^2a_f^2}{m^2}\partial_{i_1}\partial_{i_2}\int\int d\eta_1 d\eta_2 a^{-2}(\eta_1)a^{-2}(\eta_2)\langle\phi(\eta_1, r_1)\phi(\eta_2, r_2)\rangle_{flat}.
\end{equation}

Note that Eq. (\ref{eqproperv1}) is the proper velocity correlation function for free particles and Eq. (\ref{eqproperv2}) is the proper velocity correlation function for bounded particles. In the latter, the scale factor appears in the integrand of Eq. (\ref{eqproperv2}). We will see that the result found in this case is different from the free case analyzed in Sect. \ref{secfrw}. The first difference appears when we consider only the pure Minkowski contribution from Eq. (\ref{eqboundary1}) or in other words, when no planes are present. The result that will be obtained for the correlation function is different from zero.  

Now we can specify the kind of expansion by defining the scale factor ($a = a_{eff}$). Let us consider an asymptotically bound dimensionless scale factor given by 

\begin{equation}\label{eqtanh}
a^2(\eta) = a_0^2 + a_1^2\tanh(\eta/\eta_0).
\end{equation}
This scale factor is considered in many analog models for a FRW universe (see Refs. \cite{jwvg07,pfl10}). The time $\eta_0 $ is constant and it determines the rate of the expansion. Note that when $\eta \rightarrow \pm \infty$, $a^2 \rightarrow a_0^2 \pm a_1^2$, so a realistic scale function implies $a_0 > a_1$. We can see from Eq. (\ref{eqtanh}) that the constants $a_0$ and $a_1$ are dimensionless parameters and can be defined in terms of  

\begin{equation}\label{eqa0}
a_0^2 = \frac{a_f^2 + a_i^2}{2}
\end{equation}
and

\begin{equation}\label{eqa1}
a_1^2 = \frac{a_f^2 - a_i^2}{2} 
\end{equation}
where $a_i$ is the initial scale factor, i.e. when the expansion begins, and $a_f$ is the final one. They can be written  in terms of the fluid parameters defined in Section \ref{secbackground}. Note that $a_i^2 = a_{eff}^2(\eta = \eta_i) = 1$ once we have chosen  $b(t_0) = 1$. In fact, we will choose the scale factor given by Eq. (\ref{eqtanh}) to evaluate the dispersion in the velocity of the particles. In this case the scalar field correlation function to this model is given by Eq. (\ref{eqconforme1}) with Eqs. (\ref{eqaeff77}) and (\ref{eqtanh}).

\subsubsection{First case: no plane boundary}\label{seczero}

Let us consider the case with no plane boundaries  by introducing the first term of Eq. (\ref{eqboundary1}) into Eq. (\ref{eqproperv2}) and particularizing our study to the $z$ direction. The results in the $x$ and $y$ directions are identical due to the fact that we are dealing with an homogeneous and isotropic background geometry and no plane boundary is present. Thus,  after the derivations with respect to $z_1$ and $z_2$, Eq. (\ref{eqproperv2}) reads

\begin{eqnarray}\label{noplate1}
\langle v^z(\eta_1, r_1)v^z(\eta_2, r_2)\rangle_0 = \frac{q^2a_f^2}{2\pi^2m^2}\int_{-\infty}^{+\infty}\int_{-\infty}^{+\infty} d\eta_1 d\eta_2 a^{-2}(\eta_1)a^{-2}(\eta_2)\{f_2(\eta, z)  \\ \nonumber  + 4\Delta z^2f_3(\eta, z)\},
\end{eqnarray}
where the function $f_n(\eta, z)$ is given by

\begin{eqnarray}
f_n(\eta, z) = \frac{1}{[c_0^2(\eta_1-\eta_2)^2 - r^2]^n},
\end{eqnarray}
with $r$ being the same defined in Eq. (\ref{eqr}). The subscript ($\langle \;\; \rangle_0$) included in the left hand side of Eq. (\ref{noplate1}) indicates that no plane boundaries are present.

The solution of the integral in Eq. (\ref{noplate1}) is

\begin{eqnarray}
\langle v^z(\eta_1, r_1)v^z(\eta_2, r_2)\rangle_0 = \frac{2q^2a_f^2A}{\pi^4m^2c_0^4\eta_0^2a_1^4}\left\{ S_2(r) - 4\frac{\Delta z^2}{\pi^2c_0^2\eta_0^2} S_3(r) \right\},
\end{eqnarray}
where 

\begin{equation}\label{sumsn}
S_n(r) =\sum_{p=2}^{\infty}\frac{(p-1)}{[p^2+b^2]^n},
\end{equation}
with

\begin{equation}
b^2  = \frac{r^2}{\pi^2c_0^2\eta_0^2}.
\end{equation}
and $A$ is a constant of integration defined by $A \equiv \sinh^4\left[\frac{1}{2}\ln\left(\frac{a_0^2 + a_1^2}{a_0^2 - a_1^2}\right)\right]$. To evaluate the integrals given in Eq. (\ref{noplate1}) we have followed the same steps of Appendix 2 of Ref. \cite{bbf09}.

The sums $S_2$ and $S_3$ are show in Appendix \ref{ap1} of the present paper in Eqs. (\ref{eqp1}) and (\ref{eqp2}). Taking the coincidence limit in those equations, we obtain 

\begin{eqnarray}\label{vvz1}
\langle (\Delta v^z)^2\rangle_0 = \frac{2q^2B}{\pi^4m^2c_0^4\eta_0^2}\left(\zeta(3) - \frac{\pi^4}{90}\right),
\end{eqnarray}
in terms of a zeta function ($\zeta(3)$) and $B \equiv a^2_fA/a^4_1$, a dimensionless constant that can be written in terms of the fluid parameters by

\begin{equation}\label{eqB}
B = \frac{c_0}{4c_{sf}} \left(1 - \frac{c_{sf}}{c_0}\right)^2
\end{equation}
with $c_{sf} = c_s(\eta = \eta_f)$. One can see that for an expansion, $b(t)$ decreases, and from Eq. (\ref{eqb1}) $c_{sf} < c_0$ . Note that initially $c_{sf} = c_0$, and a zero dispersion is found, which means, there is no Brownian motion. 

Differently from the free particle case, we obtain a non-zero and constant result in Eq. (\ref{vvz1}). A similar result was obtained for electric charged particles in Ref. \cite{bbf09}. Thus, for bound particles the dispersion is different from zero. One possible interpretation to this result is that now the particles are not, in the average, moving on geodesics and the expansion prevent them to move apart one to the other.

\subsubsection{Second case: one plane boundary}\label{secone}

In order to investigate the influence of a single plane boundary in Eq. (\ref{eqproperv2}) we must include in this equation the second term of Eq. (\ref{eqboundary1}). Once the contribution from a pure Minkowski term, given by the first term of (\ref{eqboundary1}) was evaluated in the Section \ref{seczero}, we will evaluate here only the contribution due the plane located at $z = 0$. Thus, in this section the total contribution to the dispersion in the velocity is the result found here plus the expression found in Eq. (\ref{vvz1}). 

To begin with, let us note that we have an expression similar to the one found in Eq. (\ref{noplate1}), which is written as

\begin{eqnarray}\label{1plate1}
\langle v^z(\eta_1, r_1)v^z(\eta_2, r_2)\rangle_1 = \frac{q^2a_f^2}{2\pi^2m^2}\int_{-\infty}^{+\infty}\int_{-\infty}^{+\infty} d\eta_1 d\eta_2 a^{-2}(\eta_1)a^{-2}(\eta_2)\{f_2(\eta, \tilde{r})  \\ \nonumber + 4\Delta \tilde{z}^2f_3(\eta, \tilde{r})\}
\end{eqnarray}
for the perpendicular direction and

\begin{eqnarray}\label{1platex1}
\langle v^x(\eta_1, r_1)v^x(\eta_2, r_2)\rangle_1 = \frac{-q^2a_f^2}{2\pi^2m^2}\int_{-\infty}^{+\infty}\int_{-\infty}^{+\infty} d\eta_1 d\eta_2 a^{-2}(\eta_1)a^{-2}(\eta_2)\{f_2(\eta, \tilde{r})  \\ \nonumber + 4\Delta x^2f_3(\eta, \tilde{r})\}
\end{eqnarray}
for the parallel direction.

  The subscript ($\langle\;\;\rangle_1$) indicates that we are dealing with a single plane boundary and, as noted before, only the second term on the right hand side of Eq. (\ref{eqboundary1}) is present.

In this case, $f_n(\eta, \tilde{r}) $ is defined by

\begin{eqnarray}
f_n(\eta, \tilde{r}) = \frac{1}{[c_0^2(\eta_1-\eta_2)^2 - \tilde{r}^2]^n},
\end{eqnarray}
with $\tilde{r}$ given by Eq. (\ref{eqrtilde}).

To integrate Eq. (\ref{1plate1}) we followed the same steps of the no plane boundary case. The result is given by

\begin{eqnarray}
\label{dispersion1}
\langle v^z(\eta_1, r_1)v^z(\eta_2, r_2)\rangle_1 = \frac{2q^2B}{\pi^4m^2c_0^4\eta_0^2}\left\{ S_2(\tilde{r}) - 4\frac{\Delta \tilde{z}^2}{\pi^2c_0^2\eta_0^2} S_3(\tilde{r}) \right\},
\end{eqnarray}
and

\begin{eqnarray}
\label{dispersion2}
\langle v^x(\eta_1, r_1)v^x(\eta_2, r_2)\rangle_1 = \frac{-2q^2B}{\pi^4m^2c_0^4\eta_0^2}\left\{ S_2(\tilde{r}) - 4\frac{\Delta x^2}{\pi^2c_0^2\eta_0^2} S_3(\tilde{r}) \right\},
\end{eqnarray}
where 

\begin{equation}
S_n(\tilde{r}) =\sum_{p=2}^{\infty}\frac{(p-1)}{[p^2+\tilde{b}^2]^n},
\end{equation}
with
\begin{equation}
\tilde{b}^2  = \frac{\tilde{r}^2}{\pi^2\eta_0^2c_0^2}.
\end{equation}

Note that  in the limit of coincidence, $\tilde{b}^2 = 4z^2/\pi^2c_0^2\eta_0^2$. We see from this expression that when the particles are located at the plane position, i. e. $z = 0$, the sum above can be given in terms of a zeta function and always converges for any integer number ($n$) bigger than one ($n > 1$). The general expression for $S_2(\tilde{r})$ and $S_3(\tilde{r})$ are given by Eqs. (\ref{eqp1}) and (\ref{eqp2}), respectively, when we make the substitution of $b$ by $\tilde{b}$ in those equations.    


\begin{figure}[htbp]
	\centering
		\includegraphics[scale=0.5]{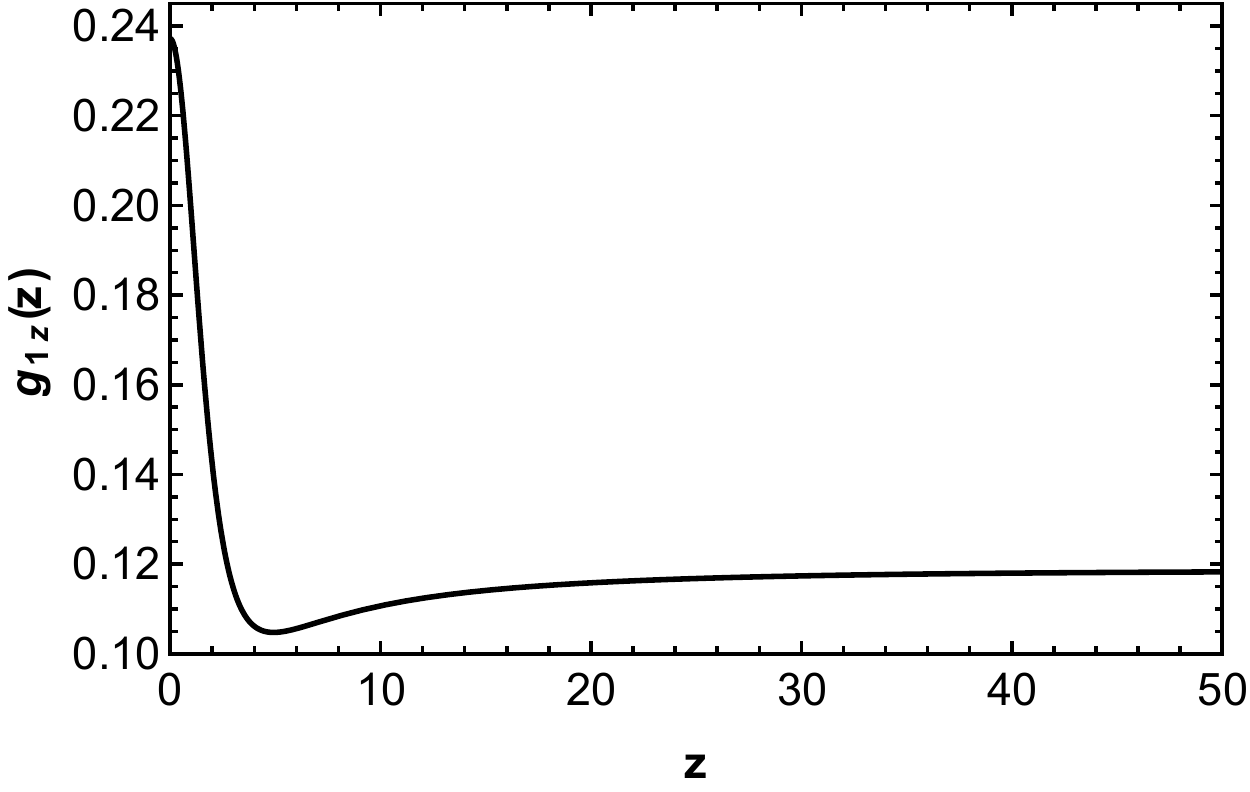}
		\caption{ {\it This figure was made in units of $g_{1z}(z) \equiv \langle(\Delta v^z)^2_{total1}\rangle m^2\eta_0^2c_0^4/Bq^2$. Perpendicular quantum dispersion for bound particle velocity in terms of $z$  has no singular points. The effect is bigger close to the plane at $z = 0$ and remains constant far from it due the time-dependent background geometry .}}
	\label{fig1planez}
\end{figure}

\begin{figure}[htbp]
	\centering
		\includegraphics[scale=0.5]{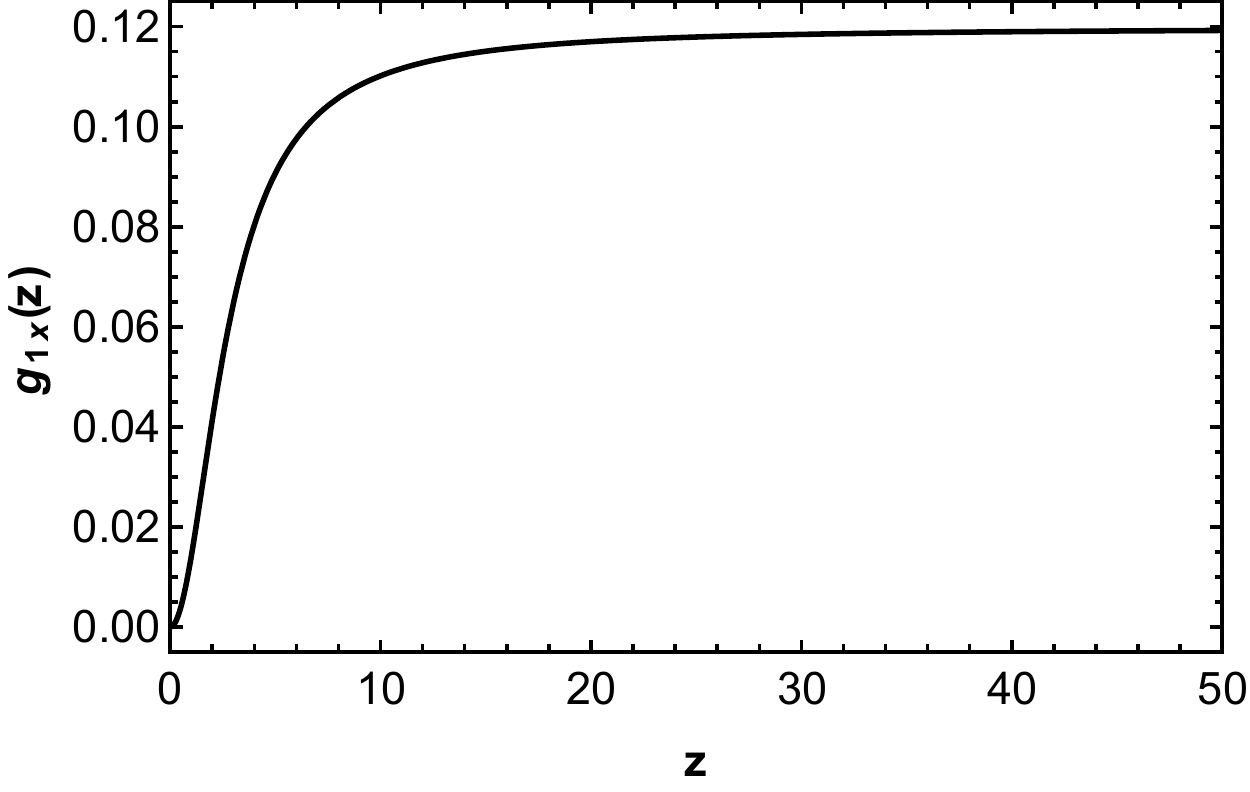}
		\caption{{\it This figure was made in units of $g_{1x}(z) \equiv \langle(\Delta v^x)^2_{total1}\rangle m^2\eta_0^2c_0^4/Bq^2$. Parallel quantum dispersion for bound particle velocity in terms of $z$  has no singular points. The effect is smaller close to the plane at $z = 0$ and remains constant far from it due the time-dependent background geometry .}}
	\label{fig1planex}
\end{figure}

Now let us discuss the two limiting cases. First consider the case when  $0 < \tilde{b} \ll 1$, which is equivalent to assume that the particle is very close to the plane located at $z = 0$. The sums above are given by the following expressions:

\begin{equation}
S_2(\tilde{z}) \approx \left(\zeta(3) - \frac{\pi^4}{90}\right) + \left(\frac{2\pi^6}{945} - 2\zeta(5)\right)\tilde{b}^2
\end{equation}
and
\begin{equation}
S_3(\tilde{z}) \approx \left( \zeta(5) - \frac{2\pi^6}{945}\right).
\end{equation}

Then, the dispersion in the velocity close to the plane in the $z$ direction is 

\begin{eqnarray}\label{vvz2}
\langle (\Delta v^z)^2\rangle_1 = \frac{2q^2B}{\pi^4m^2c_0^4\eta_0^2}\left[\zeta(3) - \frac{\pi^4}{90} + \frac{24z^2}{\pi^2c_0^2\eta_0^2}\left(\frac{\pi^6}{945} - \zeta(5)\right)\right]
\end{eqnarray}
The total velocity dispersion in $z$ direction is given by both terms of Eq. (\ref{eqboundary1}). Thus $\langle (\Delta v^z)^2\rangle_{total1} = \langle (\Delta v^z)^2\rangle_0 + \langle (\Delta v^z)^2\rangle_1$ and the total dispersion is 

\begin{eqnarray}\label{vvztotal2}
\langle (\Delta v^z)^2\rangle_{total1} = \frac{2q^2B}{\pi^4m^2c_0^4\eta_0^2}\left[2\zeta(3) - \frac{\pi^4}{45} + \frac{24z^2}{\pi^2c_0^2\eta_0^2}\left(\frac{\pi^6}{945} - \zeta(5)\right)\right].
\end{eqnarray}
When the particle is on the plane, $z = 0$, we obtain a regular result which is twice the result corresponding to the absence of planes.

The result to the $x (y)$ direction (the perpendicular direction) is, 

\begin{eqnarray}\label{vvxtotal2}
\langle (\Delta v^x)^2\rangle_{total1} = \langle (\Delta v^y)^2\rangle_{total1} =\frac{8q^2Bz^2}{\pi^4m^2c_0^6\eta_0^4}\left(\zeta(5) - \frac{\pi^6}{945}  \right).
\end{eqnarray}
Note that on the plane the dispersion is zero and the term on the left hand side is positive for a $z \neq 0$.

Another interesting limit case is when $\tilde{b} \gg 1$, which is valid for $z \gg c_0\eta_0$. In this case

\begin{eqnarray}\label{vvz3}
\langle (\Delta v^z)^2\rangle_1 = -\langle (\Delta v^x)^2\rangle_1 = -\langle (\Delta v^y)^2\rangle_1 = \frac{-q^2B}{4\pi^2m^2c_0^2z^2},
\end{eqnarray}
and the total dispersion is

\begin{eqnarray}\label{vvztotal3}
\langle (\Delta v^z)^2\rangle_{total1} = \frac{2q^2B}{\pi^4m^2\eta_0^2c_0^4}\left[\zeta(3) - \frac{\pi^4}{90} - \frac{\pi^2\eta_0^2c_0^2}{8z^2}\right],
\end{eqnarray}
for perpendicular and

\begin{eqnarray}\label{vvxtotal3}
\langle (\Delta v^x)^2\rangle_1 = \langle (\Delta v^y)^2\rangle_1 = \frac{2q^2B}{\pi^4m^2\eta_0^2c_0^4}\left[\zeta(3) - \frac{\pi^4}{90} + \frac{\pi^2\eta_0^2c_0^2}{8z^2}\right]
\end{eqnarray}
for parallel directions.

Therefore for large values of $z$, the particle is far from the plane, $\langle (\Delta v^z)^2\rangle_{total1} \approx \langle (\Delta v^z)^2\rangle_{0}$. As expected, the same result is found to the $x(y)$ directions in this limit. The results discussed in this Section is shown in Figs. \ref{fig1planez} and \ref{fig1planex}. These figures was made in units of $g_{1z}(z) \equiv \langle(\Delta v^z)^2_{total1}\rangle m^2\eta_0^2c_0^4\pi^4/2Bq^2$. In the $z$ direction the effect is bigger close to the plane however in $x(y)$ direction it is null on the plane. This results shows that the Brownian motion of test scalar bound  particles in an analog FRW geometry, in the presence of one plane boundary and subject to scalar quantum vacuum fluctuations presents an anisotropy close to the boundary and a non-null isotropy far from it.  

Note that in all cases studied here no singularity due the particle position or the flight time  were found. This is a consequence of the scale factor given by Eq. (\ref{eqtanh}). In some sense, the consequences of the time dependent background with the scale factor chosen in this paper is similar to a switching function which regularizes integrals like the one considered here. A detailed study of these functions was considered in Refs. \cite{lms14} and \cite{lrs16}  for a flat spacetime. 

For the Neumann boundary condition the total velocity dispersions can be obtained from \eqref{dispersion1} and \eqref{dispersion2}, by changing their corresponding signal. It is worth calling attention to the fact that in this case, the dispersion in the component of the velocity perpendicular to the plane at $z=0$ is smaller close to it. Otherwise, the dispersion of the parallel component of the velocity is bigger close to the plane. These behaviours are opposite to the corresponding ones obtained when Dirichlet boundary conditions are used.

\subsubsection{Third case: two planes boundaries}\label{sectwo}

Now we want to study the effects of two planes on the motion of scalar particles in the analog FRW geometry. Let us start introducing the two last terms on the brackets of the two point function given by Eq. (\ref{eqboundary2}) in the integrand of Eq. (\ref{eqproperv2}). Thus, we need to evaluate the following integrals in the $z$-direction 

\begin{figure}[htbp]
	\centering
		\includegraphics[scale=0.5]{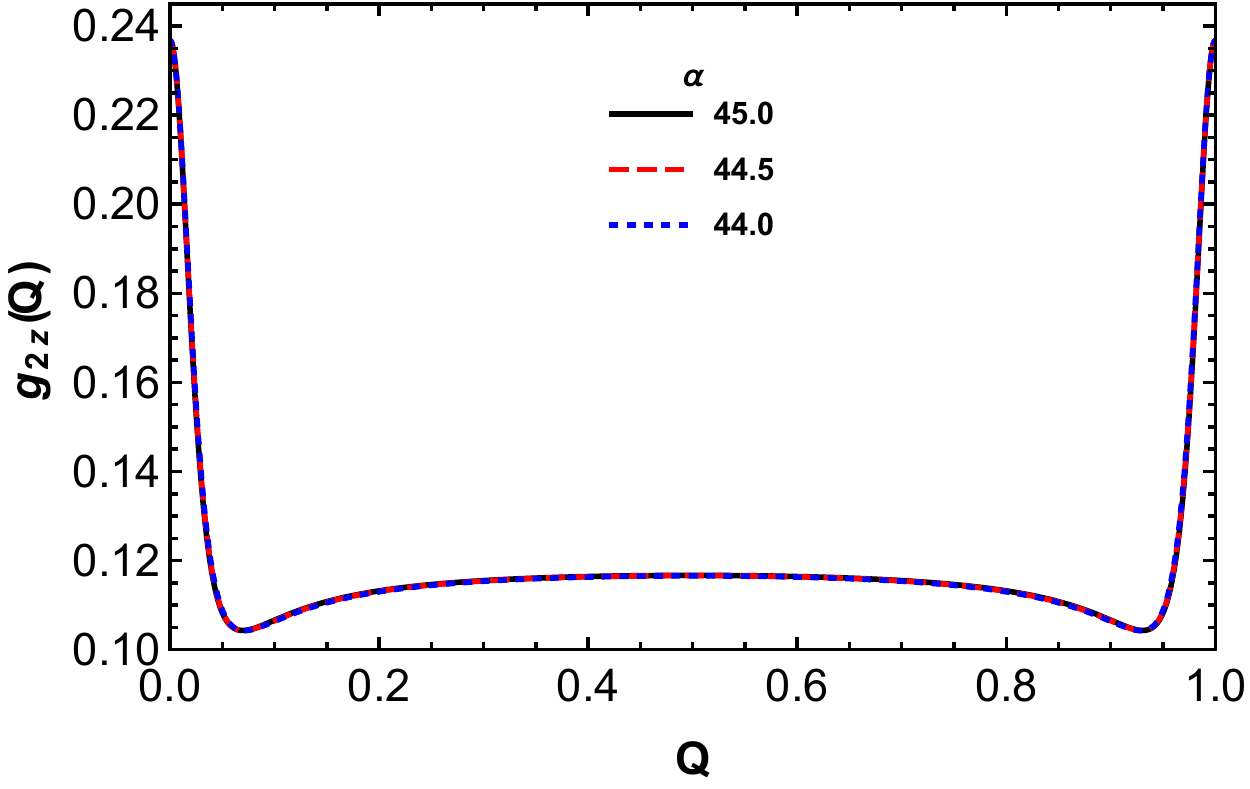}
		\caption{ {\it This figure was made in units of $g_{2z}(Q) \equiv \langle(\Delta v^z)^2_{total2}\rangle m^2\eta_0^2c_0^4/Bq^2$ and $\alpha = 2d/\pi \eta_0 c_0$. Perpendicular quantum dispersion for bound particle velocity in terms of $Q$  has no singular points. The effect is bigger close to the planes at $Q = 0$ and $Q = 1$ and remains constant in the region between the planes due the time-dependent background geometry. As the planes are far from each other, the dispersion behaves like two independent planes and does not depend on the values of $\alpha$.}}
	\label{fig2planez}
\end{figure}

\begin{figure}[htbp]
	\centering
		\includegraphics[scale=0.5]{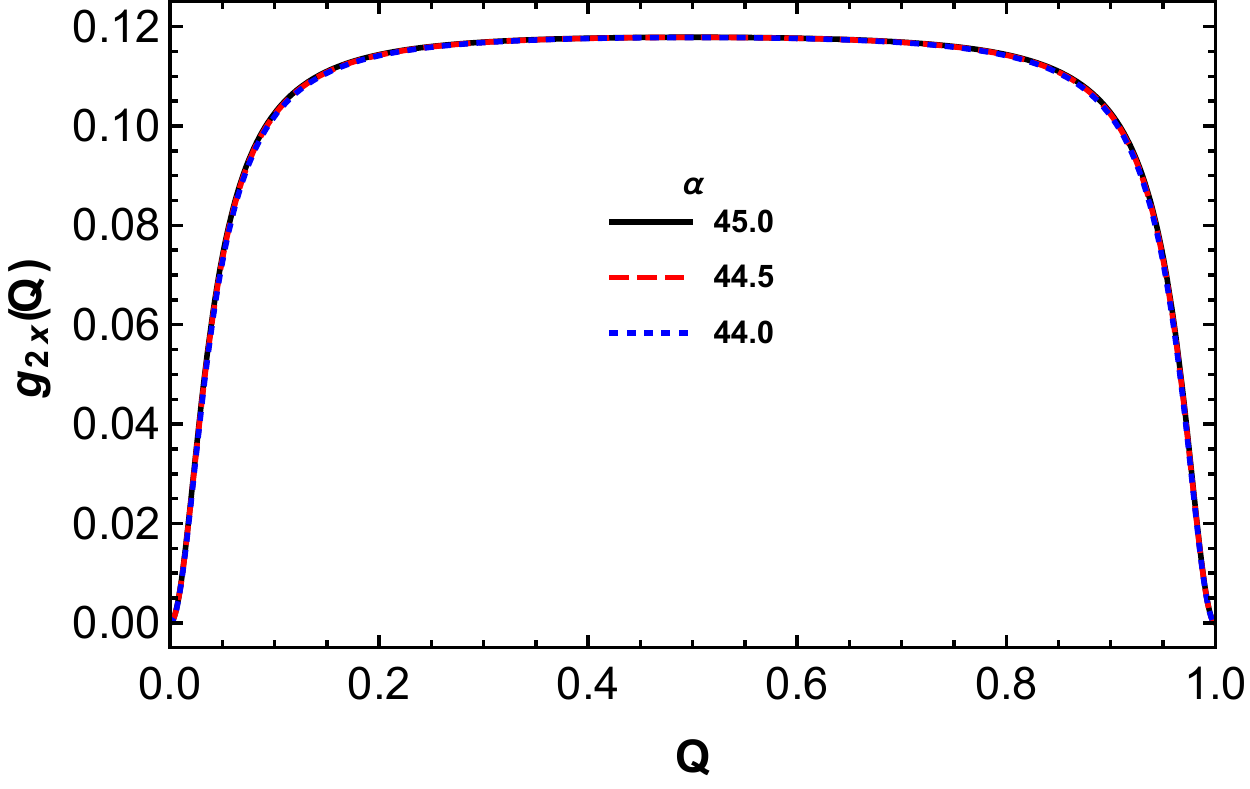}
		\caption{{\it This figure was made in units of $g_{2x}(Q) \equiv \langle(\Delta v^x)^2_{total2}\rangle m^2\eta_0^2c_0^4/Bq^2$ and $\alpha = 2d/\pi \eta_0 c_0$. Parallel quantum dispersion for bound particle velocity in terms of $Q$  has no singular points. The effect is smaller close to the planes at $Q = 0$ and $Q = 1$ and remains constant in the region between the planes due the time-dependent background geometry. As the planes are far from each other, the dispersion behaves like two independent planes and does not depend on the values of $\alpha$}}
	\label{fig2planex}
\end{figure}

\begin{figure}[htbp]
	\centering
		\includegraphics[scale=0.5]{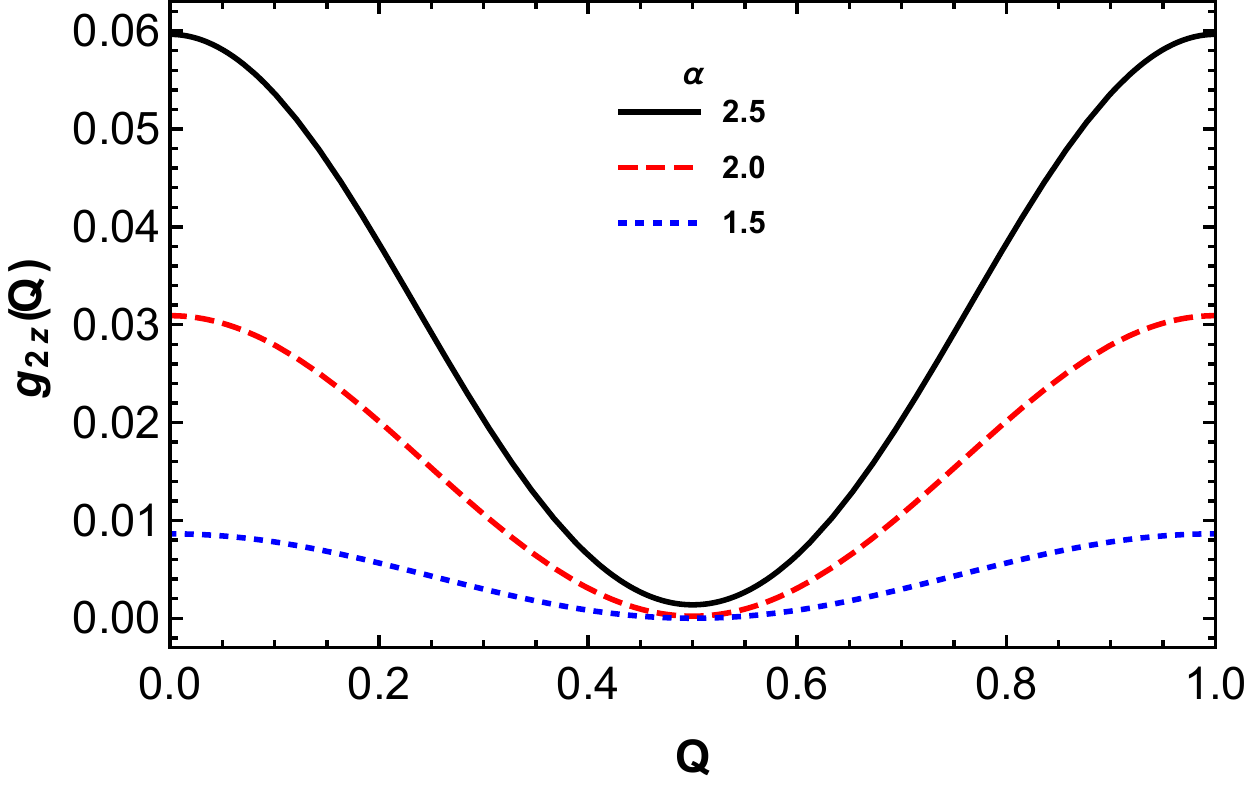}
		\caption{{\it This figure was made in units of $g_{2z}(Q) \equiv \langle(\Delta v^z)^2_{total2}\rangle m^2\eta_0^2c_0^4/Bq^2$ and $\alpha = 2d/\pi \eta_0 c_0$. Perpendicular quantum dispersion for bound particle velocity in terms of $Q$  has no singular points. The effect is bigger close to the planes at $Q = 0$ and $Q = 1$ and decrease to a non-null minimum value in the region between the planes due the time-dependent background geometry. As the planes are close to each other both  act on the particles in the whole region between them. In this situation, the dispersion increases with $\alpha$}}
	\label{fig2planez2}
\end{figure}

\begin{figure}[htbp]
	\centering
		\includegraphics[scale=0.5]{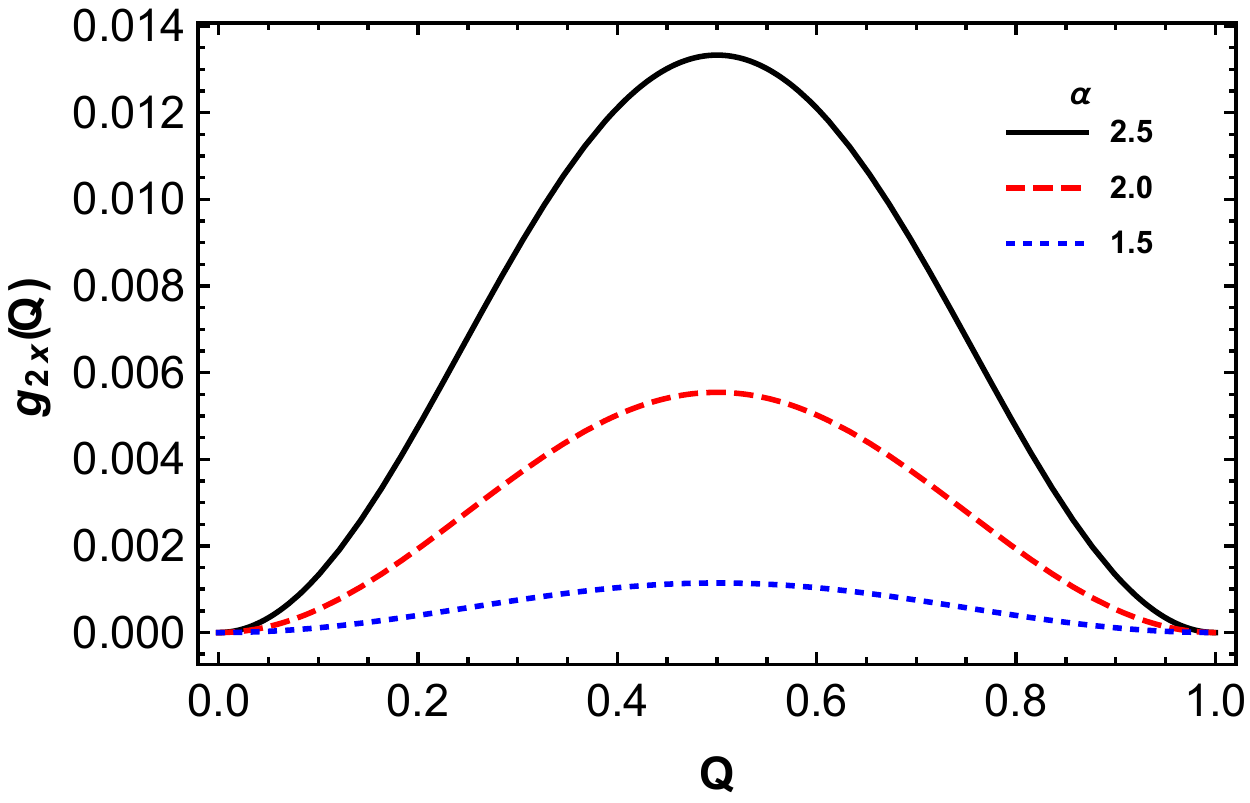}
		\caption{{\it This figure was made in units of $g_{2x}(Q) \equiv \langle(\Delta v^x)^2_{total2}\rangle m^2\eta_0^2c_0^4/Bq^2$ and $\alpha = 2d/\pi \eta_0 c_0$. Parallel quantum dispersion for bound particle velocity in terms of $Q$  has no singular points. The effect is smaller close to the planes at $Q = 0$ and $Q = 1$ and increase to a maximum value in the region between the planes due the time-dependent background geometry. As the planes are close to each other both  act  on the particles in the whole region between them. In this situatio, the dispersion increases with $\alpha$.  }}
	\label{fig2planex2}
\end{figure}

\begin{eqnarray}
\langle v^z(\eta_1, r_1)v^z(\eta_2, r_2)\rangle_2 = \frac{q^2a_f^2}{2\pi^2m^2}\int_{-\infty}^{+\infty}\int_{-\infty}^{+\infty} d\eta_1 d\eta_2 a^{-2}(\eta_1)a^{-2}(\eta_2)\left \{f_2(\eta, r_l) +  \right. \\ \left. \nonumber  4\Delta z_l^2f_3(\eta, r_l) +f_2(\eta, \tilde{r}_l) + 4\Delta \tilde{z}_l^2f_3(\eta, \tilde{r}_l)\right \},
\end{eqnarray}
and in a similar way in the $x(y)$-direction.

\begin{eqnarray}
\langle v^x(\eta_1, r_1)v^x(\eta_2, r_2)\rangle_2 = \frac{q^2a_f^2}{2\pi^2m^2}\int_{-\infty}^{+\infty}\int_{-\infty}^{+\infty} d\eta_1 d\eta_2 a^{-2}(\eta_1)a^{-2}(\eta_2)\left \{f_2(\eta, r_l) +  \right. \\ \left. \nonumber  4\Delta x^2f_3(\eta, r_l) - f_2(\eta, \tilde{r}_l) - 4\Delta x^2f_3(\eta, \tilde{r}_l)\right \},
\end{eqnarray}
where the subscript ($\langle\;\;\rangle_2$) in both cases  indicates that we are dealing with two planes. 

The functions $f_n$ are defined by

\begin{eqnarray}
f_n(\eta, r_l) = \frac{1}{[c_0^2(\eta_1-\eta_2)^2 - r_l^2]^n}
\end{eqnarray}
and

\begin{eqnarray}
f_n(\eta, \tilde{r}_l) = \frac{1}{[c_0^2(\eta_1-\eta_2)^2 - \tilde{r}_l^2]^n}.
\end{eqnarray}

We can calculate the integrals above following the same steps of previous sections, which gives us the result

\begin{eqnarray}\label{eqtwo1}
\langle v^z(\eta_1, r_1)v^z(\eta_2, r_2)\rangle_2 = \frac{2q^2B}{\pi^4m^2\eta_0^2c_0^4}\left\{{{\sum}_{l=-\infty}^{'\infty}}\left[ S_2(r_l) - \frac{4\Delta z_l^2}{\pi^2\eta_0^2c_0^2}S_3(r_l)\right] + \right. \\ \nonumber \left. {\sum}_{l=-\infty}^\infty\left[ S_2(\tilde{r}_l) - \frac{4\Delta\tilde{z}_l^2}{\pi^2\eta_0^2c_0^2}S_3(\tilde{r}_l) \right]\right\},
\end{eqnarray}
and

\begin{eqnarray}\label{eqtwox1}
\langle v^x(\eta_1, r_1)v^x(\eta_2, r_2)\rangle_2 = \frac{2q^2B}{\pi^4m^2\eta_0^2c_0^4}\left\{{{\sum}_{l=-\infty}^{'\infty}}\left[ S_2(r_l) - \frac{4\Delta x^2}{\pi^2\eta_0^2c_0^2}S_3(r_l)\right] + \right. \\ \nonumber \left. - {\sum}_{l=-\infty}^\infty\left[ S_2(\tilde{r}_l) - \frac{4\Delta x^2}{\pi^2\eta_0^2c_0^2}S_3(\tilde{r}_l) \right]\right\},
\end{eqnarray}
where 

\begin{equation}\label{eqs1notildel}
S_n(r_l) =\sum_{p=2}^{\infty}\frac{(p-1)}{[p^2+b_l^2]^n},
\end{equation}
with


\begin{equation}\label{bl2p}
b_l^2  = \frac{r_l^2}{\pi^2\eta_0^2c_0^2}.
\end{equation}
The sum given by Eq. (\ref{eqs1notildel}), and by the straightforward substitution of $r_l \rightarrow \tilde{r}_l$, is regular for any integer value of $n > 1$.


Taking the limit of coincidence in the expressions above we obtain ${b}_l^2=\frac{l^2}{T_0^2}$ and $\tilde{b}_l^2=\frac{(Q+l)^2}{T_0^2}$, while $\Delta z^2 = 4d^2l^2$ and $\Delta \tilde{z}^2_l = 4d^2(Q+l)^2$. We have defined $Q = z/d$ and $T_0 = \pi\eta_0c_0/2d$. Note that $b_l$ and $\tilde{b}_l$ depend on all values of $l$ and the expansion made in the previous section for big and small values of $b$ is not trivial here. Moreover now we have a double sum, one for $p$  and one for $l$, in each term of Eqs. (\ref{eqtwo1}) and (\ref{eqtwox1}). Thus, we are unable to give a close expression, like the ones given in Eqs. (\ref{eqp1}) and (\ref{eqp2}). In this case we can study numerically the behavior of Eqs. (\ref{eqtwo1}) and (\ref{eqtwox1}) in the region between the planes. In this way, it is more convenient to make the sum over $l$ first in Eq. (\ref{eqs1notildel}) with $r_l$ or $\tilde{r}_l$ and them study numerically the sum over $p$. We show the sum over $l$ explicitly in Eqs. (\ref{eql1}), (\ref{eql2}), (\ref{eql3}) and (\ref{eql4})  in the Appendix \ref{ap1}.  

Figures \ref{fig2planez} and \ref{fig2planex}  show, respectively, the behavior of the total perpendicular and parallel dispersion of the particles in the region between the planes. In these figures the planes are far from each other (in next section we discuss what we mean by far or close planes) and different values of $\alpha$ are taking into account. Note that, comparing Figs. \ref{fig2planez} and \ref{fig2planex} with Figs. \ref{fig1planez} and \ref{fig1planex}, we can see that these dispersions behaves like as two independent planes were present. The left plane, located at $z = 0$ contributes  to the  Brownian motion of the particles close to $z = 0$, while the right plane, located at z = $d$, contributes to the particles close to $z = d$, as we should expect.  On the other hand, when the planes are close to each other, see Figs. \ref{fig2planez2} and \ref{fig2planex2}, we can see the influence of both planes under the Brownian motion of the particles. In this situation, the Brownian motion increases with $\alpha$ (we postpone to next section the discussion about the meaning of this result). In both cases we have anisotropic quantum fluctuations, the effect is bigger close to the planes for the perpendicular dispersion and smaller for parallel dispersion.    

If we consider Neumann boundary condition the total velocity dispersions can be obtained from \eqref{eqtwo1} and \eqref{eqtwox1} by changing the sign of the corresponding second terms. In this case the dispersion in the component of the velocity perpendicular to the planes at $z=0$ and $z=d$ is smaller for points close to the planes. On the other hand, the dispersion of the parallel component of the velocity is bigger for points close to the planes. These behaviours are opposite to the corresponding ones obtained when Dirichlet boundary conditions are used. Note that in this case, the results are reinforced in comparision to those of a single plane as should be expected.

\section{Summary and Discussion}\label{secconclu}

In this paper we have studied the Brownian motion of scalar test particles coupled to a fluctuating scalar field  in an analog model, that is represented by a Friedmann-Robertson-Walker spatially flat geometry. In this scenario, we have studied two kinds of situations. One was the case of free particles, which moves in average on geodesics, and the other, the case of bounded particles.  In the latter case an external force was introduced in the equation of motion of the particles. This force cancels out locally the effects of the expansion. One effect of this cancellation is that the time dependent scale factor appears in the integrand of Eq. (\ref{eqproperv2}), which represents the quantum dispersion in the velocity for bound particles, and does not appears in Eq. (\ref{eqproperv1}), which represents the quantum dispersion in the velocity for the free particles. 

Let us first discuss the results of Section \ref{secfrw}, when free particles were considered. In this case, the integrals of Eq. (\ref{eqproperv1}) are the same as in flat spacetime. Note that to recover the flat spacetime results we just need to take $a_f = 1$ in that equation. As a consequence the results found in that section are the same as the ones for a flat spacetime. When no planes are present, the dispersion is zero. When one or two planes are present the dispersion is different from zero, but it is singular at the plane positions and at a time $\eta = 2z/c_0$, for a single plane case, and at times $\eta = \pm 2z/c_0 \pm 2dl/c_0$ for double planes boundaries, as can be seen by Eq. (\ref{eqfreez1}) and Fig. \ref{figfree}, respectively. This times corresponds to a time interval of a round trip from the particles position (with velocity $c_0$) to the plane. These behaviours are the same irrespective of the boundary condition used, if Dirichlet or Neumann boundary condition.

Perhaps the most significant result of our paper is present in Section \ref{secbound}, when bounded particles were considered. In the coincidence limit a nonzero, constant and isotropic dispersion was found for these particles when no plane is present, see Eq. (\ref{vvz1}). When one plane is present and we use Dirichlet boundary conditions, we find that our result is modified by increasing the total quantum dispersion for the perpendicular proper velocity close to the plane located at $z = 0$, see Eq. (\ref{vvztotal2}) and Fig. \ref{fig1planez}, and decreasing the total quantum dispersion for the parallel proper velocity, see Eq. (\ref{vvxtotal2}) and Fig. \ref{fig1planex}. This same effect happens to the two planes case, see Figs. \ref{fig2planez} and \ref{fig2planez2}, for a perpendicular dispersion and Figs.  \ref{fig2planex} and \ref{fig2planex2} for parallel dispersion, when Dirichlet boundary condition is taken into account. On the other hand, if we use Neumann boundary condition, we get a modification in the total quantum dispersion in the opposite direction in comparision with the ones produced when Dirichlet boundary condition is considered.  

When we think about an analog model the planes could represent a physical barrier in the experiment as for instance, the one considered in Refs. \cite{fs09l} and \cite{fs09d}. A comment about the difference in the mentioned figures is necessary here. Note that Figs. \ref{fig2planez} and \ref{fig2planex} were made in a limit where the planes are far from each other, more specifically we have considered  $d \approx 70l_0$, where $l_0 = c_0\eta_0$ is the characteristic length of the phonons. In this case, the particles dispersion behaves as two independent planes were present. Thus, the plane in the left on $z = 0$ acts only on the particles close to $z = 0$ and the plane in the right on $z =d$ acts only on the particles close to $z = d$. However, in Figs. \ref{fig2planez2} and \ref{fig2planex2}, we have considered $d \approx 3l_0$, the planes are close to each other and both act on the whole particles present in the system. Moreover, the dispersion for the case $d \approx 70l_0$ is bigger than for the case  $d \approx 3l_0$. We can see this comparing the numerical values in the vertical axis of Fig. \ref{fig2planez} with Fig. \ref{fig2planez2} or Fig. \ref{fig2planex} with Fig. \ref{fig2planex2}. One plausible explanation to this effect is that when the planes are close to each other  the free path of the particles  is smaller  when compared to the particles when the planes are far from each other. Thus, Brownian motion of the particles in the former case is limited by the distance ($d$) between the planes. This explains the reason why the dispersion is increasing with $d$ in Figs \ref{fig2planez2} and \ref{fig2planex2}. However, if we increase $d$ for values around or much bigger than $d = 70l_0$, the respective figures do not change and the system remains stable.  Another significant result we have found here is that typical divergences present in flat spacetime and in Sect. \ref{secfrw}, in both cases linked to the presence of the planes, are not present here. They are regularized by an asymptotically bounded expansion represented by the scale factor of Eq. (\ref{eqtanh}). This factor has the same role of a switching sampling function which regularizes these type of integrals (see Refs. \cite{lms14,lrs16} for this same discussion in flat spacetime).

The main purposes of this paper is the theoretical study of the quantum Brownian motion induced by a vacuum fluctuating field which is a non-trivial quantum phenomena. In Minkowski spacetime this effect does not exist but a time dependent background can give a non-zero effect. Thus, an ideal scenario where these effects could be measured would be in a non-static background. An analog model using a Bose-Einstein condensate as a laboratory for some cosmological phenomena could play this role. As several of these models has been proposed in the literature and although most of them are still idealized models, as the one proposed here, the results found in this paper could be tested experimentally in a more realistic model. One possibility to measure our results is when one consider an effective temperature attributed to the particle's Brownian motion. By the use of the velocity dispersion found in Eq. (\ref{vvz1}), for instance, one can apply the non-relativistic formula: $k_BT_{eff} = m\langle(\Delta v)^2\rangle$. This is the temperature  associated to the vacuum fluctuations in the condensate. It can be estimated in terms of the critical temperature ($T_c$) of the condensate ($k_BT_c = 3.3 \hbar^2n_0^{2/3}/m$) \cite{ps08} by

\begin{eqnarray}\label{Teff}
\frac{T_{eff}}{T_c} = 9.3\times 10^{-3} \frac{q^2}{m^2c_0^4l_0^2}Ln_0^{1/3}B\approx 1.35\times 10^{-4}\frac{q^2}{c_0^4}\left(\frac{\lambda_C}{l_0}\right)^2B.
\end{eqnarray}
In the last step we used the units $\hbar = c = 1$ (note that in this units $q$ and $c_0$ are dimensionless constants) and $0 < B < 1$ and is given by Eq. (\ref{eqB}). The characteristic length of the phonons was defined by $l_0 = c_0\eta_0$  and $\lambda_C = 1/m$ is the atom Compton wavelength. In the estimation above we have considered $n_0 = 2.9\times 10^{20}m^{-3}$, $L = 2.2 \times 10^{-9}m$ and Eq. (\ref{eqvelocity}) for the sound speed. These data can be found in Ref. \cite{pfl10} for a $1 + 1$ expanding BEC model. In order to have localized individual particles, we should work in the limit where particles creations in the condensation is small. Thus we should have $l_0 > \lambda_C$. Note that, as was explained above, the result present in Eq. (\ref{Teff}) could increase with the presence of the planes.

Since we have considered  Dirichlet boundary conditions for the planes studied in Sects. \ref{secfrw} and \ref{secbound}, another natural question to acquire is the dependence of our results with these conditions. For Neumann boundary conditions, for instance, the derivative of the field considered is null at the boundaries. Once the dispersion in the velocity is proportional to the derivatives of the field, see Eqs. (\ref{eqproperv1}) and (\ref{eqproperv2}), it is to be expected for the quantum dispersion to be smaller close to the planes in the $z$-direction and bigger to the $x(y)$-direction, which is an opposed result as compared to the Dirichlet boundary conditions. Therefore, for points close to the planes, the corresponding velocity dispersions exhibit opposite behaviours and this is the main distintion between the 
two boundary conditions. It is worth emphasize that in all cases, far from the planes the velocity dispersions remain constant. These results tell us that the planes gives an anisotropic dispersion close to the planes and an isotropic dispersion far from them, 
independently of the boundary condition under consideration.  

Finally we would like to point it out that although the measurability of the results found in this paper are not an easy task with current tecnology, this work provides an operational mean to try phenomenologically observe this type of quantum vacuum effects.


\appendix
\section{Sums}\label{ap1}

In this appendix we start showing the sums $S_2$ and $S_3$ of Eq. (\ref{sumsn}). The sum over $p$ give us the result:

\begin{eqnarray}\label{eqp1}
S_2(r) = \frac{1}{4b^4}\left\{ 2+ \pi^2b^2-\pi b\coth(\pi b) - \pi^2b^2\coth(\pi b)\right. + \\ \left. \nonumber i\left[\pi^2b^3\coth^2(\pi b)-\pi^2b^3 + b +2b^3\psi(1,bi)\right]\right\},
\end{eqnarray}

\begin{eqnarray}\label{eqp2}
S_3(r) = \frac{1}{16b^6}\left\{ 8 - 3\pi b\coth(\pi b) + 3\pi^2b^2 + 2b^4\psi(2, bi) - 3\pi^2b^2\coth^2(\pi b) \right. \\ \nonumber  + 2\pi^3b^3\coth(\pi b)-2\pi^3b^3\coth^3(\pi b)  + \\  \nonumber i\left[3b + 2b^3\psi(1, bi) - \pi^2b^3 +\pi^2b^3\coth^2(\pi b) - 2\pi^3b^4\coth(\pi b) \right. \\ \nonumber \left.\left. + 2\pi^3b^4\coth^3(\pi b)\right]\right\},
\end{eqnarray}
where $\psi(n, bi)$ is the nth-derivative of the di-gamma function \cite{as72}.

Now we show the sum over $l$ of  Eq. (\ref{eqs1notildel}) when $n = 2$ and $n = 3$. They are, respectively

\begin{eqnarray}\label{eql1}
{\sum}_{l=-\infty}^{'\infty}S_2(r_l) = \sum_{p=2}^{\infty}\frac{(p-1)\pi\left[\alpha\coth\left(\frac{p\pi}{\alpha}\right) + p\pi csch^2\left(\frac{p\pi}{\alpha}\right) \right]}{2\alpha^2p^3},
\end{eqnarray}

\begin{eqnarray}\label{eql2}
{\sum}_{l=-\infty}^{'\infty}\alpha^2 l^2 S_3(r_l) = \sum_{p=2}^\infty \frac{1}{8\alpha^3p^3}\left\{\pi(p-1)csch^3\left(\frac{\pi p}{\alpha}\right)\left[-(\alpha^28\pi^2p^2)\cosh\left(\frac{p\pi}{\alpha}\right) \right. \right. + \\ \nonumber \left. \left.\alpha^2\cosh\left(\frac{3p\pi}{\alpha}\right) + 4p\alpha\pi\sinh\left(\frac{p\pi}{\alpha}\right)\right]\right\},
\end{eqnarray}
where $\alpha = 2d/\pi \eta_0 c_0$ and the sum over $l$ of Eq. (\ref{eqs1notildel}), taking the substitution $r_l$ by $\tilde{r}_l$, when $n = 2$ is 

\begin{eqnarray}\label{eql3}
{\sum}_{l=-\infty}^{\infty}S_2(\tilde{r}_l) = \sum_{p=2}^{\infty}\frac{(p-1)\pi}{2\alpha^2p^3}\left\{\frac{\alpha\sinh\left(\frac{2p\pi}{\alpha}\right)}{\cosh\left(\frac{2p\pi}{\alpha}\right) - \cos(2\pi Q)} +\right. \\ \nonumber \left. \frac{2\pi p\left[-1 + \cos(2\pi Q)\cosh\left(\frac{2p\pi}{\alpha}\right)\right]}{\left[\cos(2\pi Q) - \cosh\left(\frac{2p\pi}{\alpha}\right)\right]^2}\right\}
\end{eqnarray}
and when  $n = 3$ 

\begin{eqnarray}\label{eql4}
{\sum}_{l=-\infty}^{\infty}\alpha^2 (Q+l)^2 S_3(\tilde{r}_l) = \sum_{p=2}^\infty 4\alpha^2\left\{\frac{(p-1)\pi}{8\alpha^3p^3} \frac{\sinh\left(\frac{2p\pi}{\alpha}\right)}{\left[\cosh\left(\frac{2p\pi}{\alpha}\right) - \cos(2\pi Q)\right]}  - \right.  \\ \nonumber  \frac{(p-1)\pi^3}{4\alpha^5 p}\frac{\sinh\left(\frac{2p\pi}{\alpha}\right)\left[-3+\cos(4\pi Q) + 2\cos(2\pi Q)\cosh\left(\frac{2p\pi}{\alpha}\right)\right]}{\left[-\cos(2\pi Q)+\cosh\left(\frac{2p\pi}{\alpha}\right)\right]^3} +
\\ \nonumber \left.  \frac{(p-1)\pi^2}{4\alpha^4 p^2}\frac{\left[-1 +\cos(2\pi Q)\cosh\left(\frac{2p\pi}{\alpha}\right)\right]}{\left[\cos(2\pi Q) - \cosh\left(\frac{2p\pi}{\alpha}\right)\right]^2} \right\}.
\end{eqnarray}
. 

\begin{acknowledgments}
The authors thanks Brazilian agency CNPq (Conselho Nacional de Desenvolvimento Cient\'{\i}fico e Tecnol\'ogico - Brazil) for financial support. C.H.G.B is funded by the research project no 502029/2014-5. V.B.B is partially supported by the research project no 304553/2010-7. E.R.B.M is partially supported by the research project no 313137/2014-5. H.F.M is funded by the research project no 402056/2014-0. 
\end{acknowledgments}

\end{document}